\documentclass[11pt,twocolumn]{article}

\usepackage[table,x11names]{xcolor}
\usepackage[preprint]{acl}
\usepackage{rotating}
\def\method{our framework}
\usepackage{booktabs} % For formal tables
\usepackage{siunitx} % Provides the S column type
\usepackage{times}
\usepackage{latexsym}
\usepackage{amsmath}
\usepackage[T1]{fontenc}
\usepackage[utf8]{inputenc}
\usepackage{microtype}
\usepackage{inconsolata}
\usepackage{graphicx}
\usepackage{multirow}
\usepackage{lipsum}
\usepackage{placeins}
\usepackage{stfloats}
\usepackage{caption}
\usepackage[titletoc,title]{appendix}
\usepackage{float}
\usepackage{tocloft}

\title{KU-DMIS at EHRSQL 2024: 

Generating SQL query via question templatization in EHR}

\author{
  \textbf{Hajung Kim\textsuperscript{1*}},
  \textbf{Chanhwi Kim\textsuperscript{1*}},
  \textbf{Hoonick Lee\textsuperscript{1}},
  \textbf{Kyochul Jang\textsuperscript{1}},
  \textbf{Jiwoo Lee\textsuperscript{1}},
\\
  \textbf{Kyungjae Lee\textsuperscript{2}},
  \textbf{Gangwoo Kim\textsuperscript{1}},
  \textbf{Jaewoo Kang\textsuperscript{1,3$\dagger$}}
\\
\\
  \textsuperscript{1}Korea University,
  \textsuperscript{2}LG AI Research,
  \textsuperscript{3}AIGEN Sciences
\\
  \texttt{\{hajungk, chanhwi\_kim, hoonick,  gcj0125, hijiwoo7\}@korea.ac.kr}
\\
  \texttt{kyungjae.lee@lgresearch.ai}, \texttt{\{gangwoo\_kim, kangj\}@korea.ac.kr} 
}

\begin{document}
\maketitle

\newcommand\blfootnote[1]{%
  \begingroup
  \renewcommand\thefootnote{}\footnote{#1}%
  \addtocounter{footnote}{-1}%
  \endgroup
} 

\renewcommand{\cftsecleader}{\cftdotfill{\cftdotsep}}

\blfootnote{\textsuperscript{$*$} Equal contribution, \textsuperscript{$\dagger$} Corresponding author}

\begin{abstract}
Transforming natural language questions into SQL queries is crucial for precise data retrieval from electronic health record (EHR) databases. 
A significant challenge in this process is detecting and rejecting unanswerable questions that request information beyond the database's scope or exceed the system's capabilities.
In this paper, we introduce a novel text-to-SQL framework that robustly handles out-of-domain questions and verifies the generated queries with query execution.
Our framework begins by standardizing the structure of questions into a templated format. 
We use a powerful large language model (LLM), fine-tuned GPT-3.5 with detailed prompts involving the table schemas of the EHR database system. 
Our experimental results demonstrate the effectiveness of our framework on the EHRSQL-2024 benchmark benchmark, a shared task in the ClinicalNLP workshop. Although a straightforward fine-tuning of GPT shows promising results on the development set, it struggled with the out-of-domain questions in the test set. With our framework, we improve our system's adaptability and achieve competitive performances in the official leaderboard of the EHRSQL-2024 challenge.

\end{abstract}

\section{Introduction}
\begin{figure}[t]
    \includegraphics[width=\columnwidth]{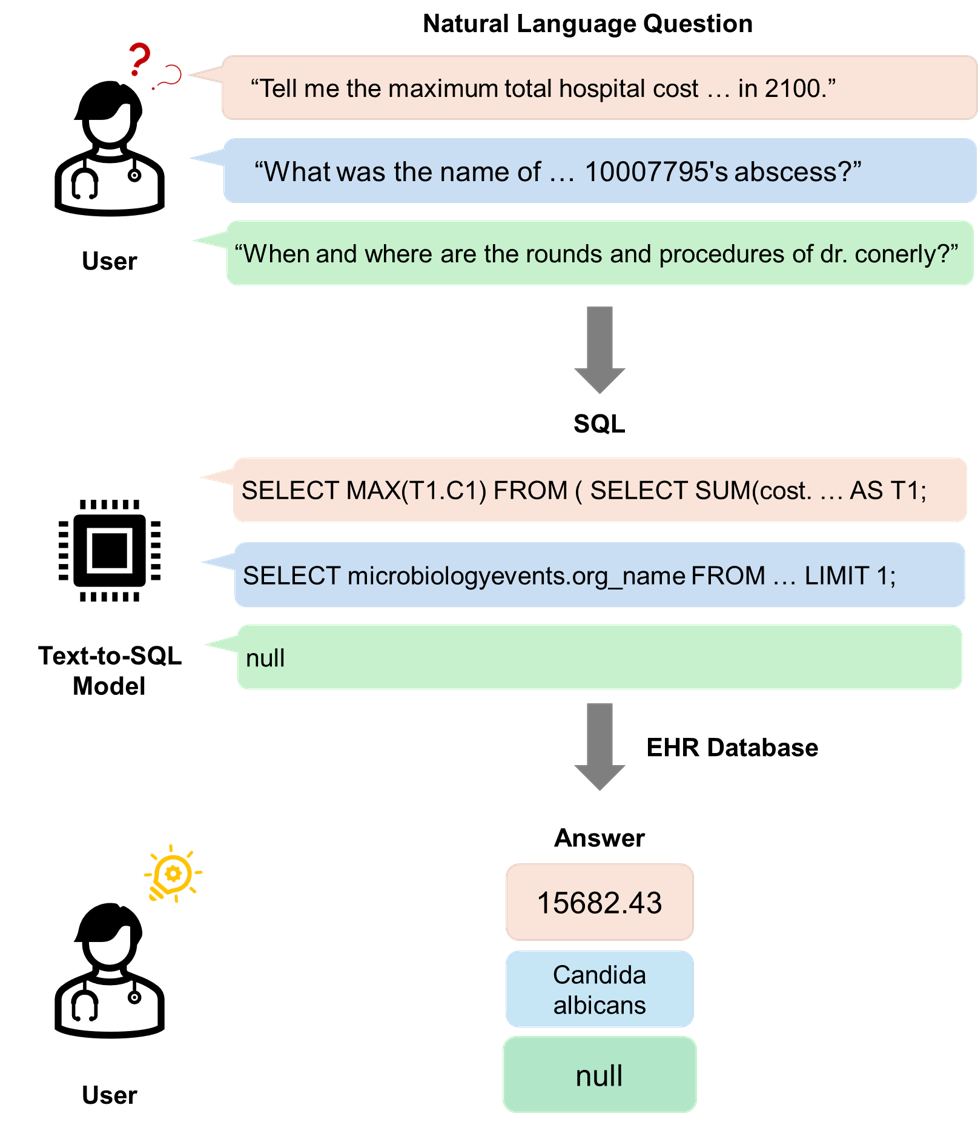}
    \caption{In the proposed Text-to-SQL framework, when a query is presented in natural language, the model generates SQL code to retrieve the required information from the database. If the query requires information absent from the database, the Text-to-SQL model returns a 'null' response.}
    \label{fig:task_def}
\end{figure}

Electronic Health Records (EHRs) are crucial elements of the contemporary healthcare system, storing patients' medical histories in relational databases. However, retrieving data from EHRs can be challenging, requiring specialized training in Structured Query Language (SQL). To bridge this gap, previous works build AI-powered systems that parse the user's question~\cite {yin2020tabert, brunner2021valuenet} or convert it into an SQL query that the database can process. \citet{lee2022ehrsql} identify an essential component in this text-to-SQL task; recognizing and adequately handling unanswerable questions that seek information beyond what the database contains. Hence, to ensure reliability and trustworthiness, systems should be able to refrain from answering unanswerable questions.

To further encourage research in this field, the Clinical NLP 2024 workshop has introduced a new shared task called EHRSQL-2024~\citep{lee2024overview} to motivate the development of more reliable question-answering (QA) systems. The EHRSQL-2024 dataset involves the real-world needs of medical personnel, incorporating templates of their most common questions. 
In this challenge, systems are tasked to generate SQL queries that accurately return the desired information from tables from the MIMIC-IV~\citep{johnson2016mimic}. Additionally, the dataset includes inherently unanswerable questions, either due to the restrictions of the database schema or the request for information not contained within the databases. On the other hand, the test set presents distracting question types that contain noisy words, further testing the robustness of participants' systems.

In this paper, we introduce a novel framework created to convert natural language questions to SQL queries for EHR databases. This framework transforms free-form questions into a templated format to handle distracting questions. We fine-tune GPT-3.5-turbo~\citep{brown2020language}, one of the most performant large language models (LLMs), optimizing it to effectively interpret intricate medical queries and produce the corresponding SQL queries. We also provide detailed prompts that describe the tables in the EHR database. 

 For SQL generation, given the task's complexity and relationships between tables, we break it down into two steps: selecting relevant tables and then generating SQL by reflecting in-depth on the selected tables. We enhance the accuracy and reliability of the generated SQL queries by correcting any errors in table names and applying ensemble techniques with majority voting.

Our empirical results of fine-tuned GPT-3.5 on the EHRSQL-2024 benchmark highlight its capability, achieving third place on the development set. However, it revealed a limitation in generalizing to questions in the test set that diverged from the predefined templates. By using our framework, we successfully address this gap between free-form questions, resulting in a notable improvement of 26.5 in the RS(10) metric in the test set. Additionally, we find that decomposing the task into two steps contributed to this success, with a significant improvement in RS (10) in the test set. 
Furthermore, by employing further verification and ensemble techniques, we attained fourth place in the EHRSQL-2024 challenge's official leaderboard.

We conduct in-depth analyses of the questions to uncover disparities in each split. In particular, we apply $N$-gram counting of the questions to highlight the distribution gaps. This variation emphasizes the need to develop a resilient model capable of adapting to and performing consistently across datasets with diverse word distributions. Additionally, we manually categorize the unanswerable questions into three distinct types.

\section{Related Works}

\subsection{Text-to-SQL Generation}

Text-to-SQL conversion requires interpreting natural language questions, matching them with the database schema, and producing accurate SQL queries that reflect the question's intent. This task is particularly challenging for individuals unfamiliar with database structures, highlighting the need for methods that translate natural language into SQL queries—a focus of ongoing research due to real-world applications. However, accurately generating SQL code from natural language is complex, mainly because of the challenges in integrating precise database knowledge into the model~\citep{Qin2022ASO, KatsogiannisMeimarakis2023ASO}.

Initially, efforts to address Text-to-SQL employed predefined rules~\citep{Sen2020ATHENA} to handle existing difficulties. The field has evolved since then to explore encoder-decoder models~\citep{Cai2017AnEF,Popescu2022AddressingLO}, and Text-to-SQL is tested on sequence-to-sequence approaches~\citep{Qi2022RASATIR}. With the rapid advancement in deep learning research, methodologies incorporating graph representation~\citep{Xu2018SQLtoTextGW, Wang2019RATSQLRS, Brock2021S2S} and attention mechanisms~\citep{Liu2023MultihopRG} have been extensively applied to Text-to-SQL tasks. Additionally, the Text-to-SQL task, tailored to real-world data, has been conducted on datasets such as WikiSQL~\citep{Zhong2017Seq2SQLGS},  Spider~\citep{Yu2018SpiderAL}, KaggleDBQA~\citep{Lee2021KaggleDBQARE}, and BIRD~\citep{Li2023CanLA}.

With the emergence of  LLMs like GPT~\citep{brown2020language} and LLaMA~\citep{Touvron2023LLaMAOA}, research leveraging these models has proliferated. Their comprehensive pretraining on massive text corpora enables them to show promising results using techniques like prompt engineering and in-context learning~\citep{Trummer2022CodexDBSC, Liu2023ACE, Chang2023HowTP, Dong2023C3ZT, Sun2023SQLPaLMIL}. Despite these advancements, exploring supervised fine-tuning has led to even greater enhancements in their performance~\citep{Gao2023TexttoSQLEB}.

\subsection{Text-to-SQL in EHR database}
The MIMIC-III~\citep{johnson2016mimic} is a prominent EHR database in the healthcare domain. MIMICSQL~\citep{Tarbell2023TowardsUT} is the first dataset constructed based on the MIMIC-III database, designing questions generated from pre-formatted templates. Similarly, emrKBQA~\citep{Raghavan2021emrKBQAAC} derived from the MIMIC-III database and the emrQA~\citep{Yue2020ClinicalRC} dataset focused on clinical reading comprehension expands the research scope. EHRSQL, introduced by \citet{lee2022ehrsql}, is an extensive text-to-SQL dataset that is linked to the two open-source EHR databases, MIMIC-III and eICU~\citep{pollard2018eicu}. Created based on feedback from 222 professionals with varied experience levels, EHRSQL covers a wide range of real-world scenarios. This dataset includes time-sensitive questions to highlight the critical importance of time in the healthcare domain. Additionally, it incorporates unanswerable questions to evaluate the system's capability to recognize and handle such inquiries effectively.

\subsection{Discriminating Unanswerable Questions}

The distinction between answerable and unanswerable questions is crucial in NLP tasks, especially in domains where accuracy and reliability are critical, such as healthcare. Discriminating between these types of questions is complex due to the subtle differences in what a question may require for a satisfactory answer. The language models often exhibit overconfidence in their ability to accurately respond to a given question. To address this, the specialized datasets have been designed through various methodologies, such as rule-based editing~\citep{Jia2017AdversarialEF}, distant supervision~\citep{Joshi2017TriviaQAAL}, and crowdsourcing~\citep{Rajpurkar2018KnowWY}, each method offering its own set of benefits and challenges for identifying unanswerable questions. This advancement facilitates more reliable and accurate question-answering capabilities, which is crucial for applications where the cost of misinformation can be high.
\section{Methods}

\begin{figure*}[t]
    \includegraphics[width=\linewidth]{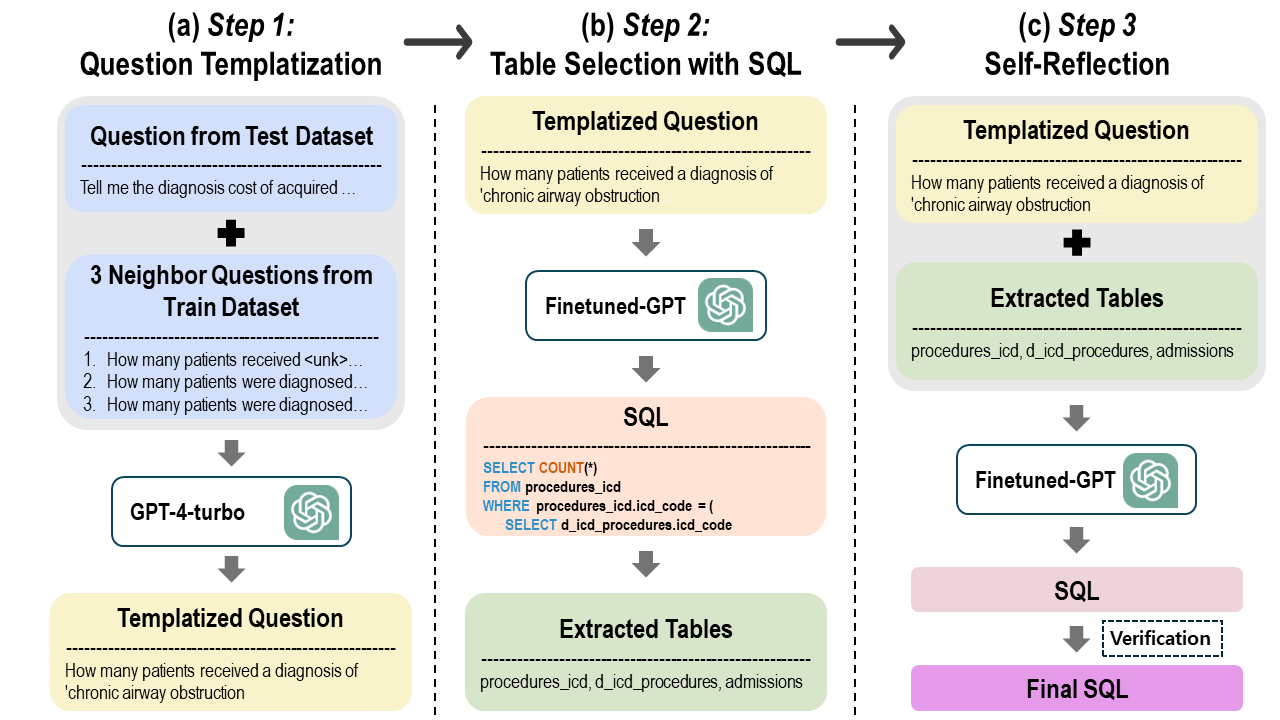}
    \caption{Overview of our framework. (a) Question Templatization (Sec.~\ref{sec:question Template}). Implementing question templatization to convert free-form questions into a structured format. (b) SQL Generation (Sec.~\ref{sec:SQL generation}). Providing task outlines and table information to aid in precise query generation. (c) Self-Reflection and Verification (Sec.~\ref{sec:SQL generation}, ~\ref{sec:SQL Verification}). Providing detailed table information identified in the initial SQL generation and then finalizing the process.}
    \label{fig:pipeline}
\end{figure*}

Figure~\ref{fig:pipeline} presents an outline of our proposed methodology. Our process starts with the templatization of questions, transforming free-form inquiries into a standardized format to ensure consistency in how queries are represented. Additionally, we enrich the model's understanding by supplying detailed information about the database tables, thereby improving its capacity to formulate precise queries. To further elevate the accuracy of the generated SQL queries, we introduce a verification phase to confirm that the queries accurately correspond to the intended data retrieval objectives. \\

\subsection{Question Templatization}
\label{sec:question Template}
We introduce question templatization to handle the diverse forms of question presentation. This approach addresses the challenge of questions deviating from a standard template by employing a reverse engineering strategy. By converting free-form questions into a templated format, we aim to align them more closely with similar patterns, thus bridging the gap between the varied question formats in real-world contexts. Specifically, we prompt GPT-4-turbo to rewrite questions to match the structure of pre-defined templates more closely.

Identifying semantically close questions involves searching for questions similar to the input question. This similarity is quantified by calculating the Euclidean distance between the question embeddings and comparing input questions to potential neighbors. We mask identification information to ensure that specific table values do not skew this comparison. For example, a question like \textit{"Count how many times in the first hospital visit patient 10004457 had coronary arteriography using two catheters."} is transformed into \textit{"Count how many times patient \textbf{<patient number>} had \textbf{<procedure>} during their first hospital visit ."} By adopting this method, we achieve a uniform question format, effectively standardizing free-form queries and reducing discrepancies in dataset distribution. The templatized question is utilized as the input question. \\

\subsection{SQL generation}
\label{sec:SQL generation}
Considering the complexity of the text-to-SQL task and the intricate relationships among more than ten tables in the database, we propose a two-stage approach that involves a table selection phase followed by a self-reflection phase. 

\paragraph{Table Selection} 
We task the GPT model with converting natural language questions into SQL queries. The construction of prompts for the model involves three essential components: (1) outlining the text-to-SQL task 
 guides the model to convert a natural language question into an SQL query for data retrieval from the EHR database. We clarify that the database uses SQLite and highlight the syntactical nuances between SQLite and other SQL dialects to guide the model's syntax choice. 
 (2) By detailing the database tables, we describe the database's complex structure, listing over ten tables with brief descriptions and their respective columns. This detail is crucial since it aids the model in identifying the relevant tables and navigating their relational schema without direct access to the database values. We follow the format introduced in DAIL-SQL~\citep{Gao2023TexttoSQLEB} for table schema details, which allows both natural language and SQL representations. 
 (3) Presenting the question for conversion is the natural language question to be transformed. 
 By using this prompt, we prompt the model to produce an SQL query that matches the question and subsequently identifies the table name mentioned within the SQL query. 

\paragraph{Self-Reflection} The prompt for the self-reflection stage is similar to Table Selection, except for detailing table information. In this stage, the prompt is augmented with detailed descriptions for each table column identified in the initial SQL query. This refinement aims to enhance the SQL query formulation by providing a more comprehensive understanding of the selected table's specifics, enabling the model to generate a more accurate and targeted SQL query.\\

\subsection{SQL Verification}
\label{sec:SQL Verification}
We implement a verification step on the generated SQL queries to address two specific scenarios. In the first scenario, some questions can be technically converted into SQL queries but remain unanswerable due to the absence of required information in the dataset. To avoid providing incorrect SQL results and improper answers, which are unanswerable, we verify the validity of each SQL query by executing it against the database. If the execution results in an error, indicating the SQL query cannot retrieve the correct answer, we replace the SQL query with \textit{null} instead. This adjustment ensures the query is considered valid but unanswerable, optimizing score outcomes.

The second scenario addresses instances where the generated SQL query includes incorrect references to table names or column names. In such cases, we identify the correct table name and associated column names based on the table values mentioned in the SQL query. We then modify the SQL query to accurately reflect the proper table name and column names to which the table values correspond. This correction process ensures that the SQL query accurately represents the intended data retrieval operation, aligning with the database's schema.\\

\subsection{Ensemble with Majority Voting}
We incorporate an ensemble method to determine the final SQL query. We first instruct GPT-4-turbo to evaluate whether the generated SQL query accurately captures the intent of the original natural language question. This alignment check ensures that the model prioritizes the core intent of the query, such as using the 'COUNT' function in SQL queries asking for a count of patients. To finalize the SQL query or its resulting answer from the database execution, we adopt a majority voting system. This ensemble strategy mitigates the variability inherent in the fine-tuned model and improves the robustness. Using majority voting to select the SQL query or derive its answer aims to improve performance metrics by effectively managing \textit{null} responses. \\

\section{Experiments}
~\begin{table*}[t]
    \centering
    \scalebox{0.78}{
\begin{tabular}{@{}lSSSSSSSSSS@{}}
\toprule
& \multicolumn{4}{c}{Development} & \multicolumn{4}{c}{Test} \\
\cmidrule(r){2-5} \cmidrule(l){6-9}
Team & {RS(0)} & {RS(5)} & {RS(10)} & {RS(N)} & {RS(0)} & {RS(5)} & {RS(10)} & {RS(N)} \\
\midrule
LG AI Research \& KAIST & 90.37 & \textbf{89.51} & \textbf{88.65} & \textbf{-109.6} & \textbf{88.17} & \textbf{84.75} & \textbf{81.32} & \textbf{-711.83} \\
PromptMind & 66.38 & 59.5 & 52.62 & -1533.62 & 82.6 & 78.75 & 74.89 & -817.4 \\
ProbGate & 84.18 & 79.45 & 74.72 & -1015.82 & 81.92 & 78.06 & 74.21 & -818.08 \\
\textbf{KU-DMIS (Ours)} & \textbf{91.57} & 82.98 & 74.38 & -1908.43 & 72.07 & 65.64 & 59.21 & -1427.93 \\
oleg1996 & 47.03 & 34.14 & 21.24 & -2952.97 & 68.89 & 56.47 & 44.04 & -2831.11 \\
LTRC-IIITH & N/A & N/A & N/A & N/A & 66.84 & 55.27 & 43.7 & -2633.16 \\
Saama Technologies & 57.78 & 50.47 & 43.16 & -1642.22 & 53.21 & 44.64 & 36.08 & -1946.79 \\
TEAM\_optimist & N/A & N/A & N/A & N/A & 14.14 & -349.61 & -713.37 & -84885.86 \\
\bottomrule
\end{tabular}
}
\begin{minipage}{\textwidth}
\captionsetup{width=\textwidth, singlelinecheck=false}
\raggedright
\caption{Official results of the leaderboard on EHRSQL-2024 dataset. The teams are ranked based on Reliability Score RS(10).}
\label{tab:leaderboard}
\end{minipage}
\end{table*}

\subsection{Experimental Setup}

\paragraph{Dataset} We evaluate \method using the EHRSQL-2024 challenge benchmark dataset ~\cite{lee2024overview}. This large-scale Text-to-SQL dataset contains 5,124 instances in the train set, 1,163 instances in the development set, and 1,167 instances in the test set, spanning 17 tables. The train dataset comprises natural language questions paired with their corresponding SQL queries. However, the SQL queries associated with the questions in the development and test sets are not provided.
\paragraph{Metric} Following \citet{lee2024overview}, we use the Reliability Score (RS). RS is unique because it rewards correct SQL queries for answerable questions ($ Q_{ans} $) and the choice to abstain from answering unanswerable questions ($ Q_{una} $). At the same time, it penalizes incorrect SQL generation for $ Q_{ans} $ and any attempt to answer $ Q_{una} $. Moreover, RS includes a penalty factor 'c' to adjust the evaluation's strictness according to specific safety requirements. The corresponding formula is as follows.

\begin{equation*}
\resizebox{0.9\linewidth}{!}{$
\begin{array}{l}
\phi_c(x) = \begin{cases}
1 & \text{if } x \in Q_{\text{ans}}, g(x)=1, \text{Acc}(x)=1 \\
0 & \text{if } x \in Q_{\text{ans}}, g(x)=0 \\
-c & \text{if } x \in Q_{\text{ans}}, g(x)=1, \text{Acc}(x)=0 \\
-c & \text{if } x \in Q_{\text{una}}, g(x)=1 \\
1 & \text{if } x \in Q_{\text{una}}, g(x)=0
\end{cases}
\end{array}
$}
\end{equation*}

\\

The adaptability of RS is demonstrated by evaluating models under four different scenarios, which vary based on the severity of the penalty term: RS(0), RS(10), and RS(N). For this challenge, the primary metric is RS(10), emphasizing the importance of accurately assessing answerable questions and refraining from generating SQL for unanswerable questions. \\

\subsection{Implementation Details}
We utilize GPT, one of the most performant Large Language Models (LLMs), to enhance the translation from text to SQL. We investigate the effectiveness of in-context learning and supervised fine-tuning methods. We employ GPT-3.5-turbo, GPT-4-turbo, and GPT-4 models for in-context learning, augmenting the prompt with three more examples. These examples consist of pairs of semantically related questions, including the input question and their corresponding SQL queries. The semantic relatedness is determined by calculating the Euclidean distance between question embeddings derived from the training dataset and the input question embedding. For supervised fine-tuning, we focus on the GPT-3.5-turbo model, the primary model available for fine-tuning. The model is prompted without including neighboring examples. Based on the evaluation results, it is clear that the supervised fine-tuning methodology is particularly effective in addressing the challenges inherent in text-to-SQL tasks. Further details are provided in section 4.4. Consequently, the fine-tuned GPT-3.5-turbo model is selected for further detailed experiments.
\\

\subsection{Leaderboard Results}
Table~\ref{tab:leaderboard} presents the scores of the participants' systems, ranked according to the RS(10) score. We secured fourth place in the test set rankings. All participating teams utilized Large Language Models (LLMs), with the top four teams, including ours, primarily employing a fine-tuned GPT model and incorporating various other techniques. This table underscores the efficacy of LLMs in addressing Text-to-SQL tasks.
\\

\begin{table}[!t]
\renewcommand{\arraystretch}{1.2}
\centering
\resizebox{\columnwidth}{!}{
\begin{tabular}{lccccc} \hline

    \textbf{Model} & \textbf{Few-shot} & \textbf{Table info.} & \textbf{RS(0)} & \textbf{RS(10)}& \textbf{RS(N)} \\ \hline\hline

GPT-3.5-turbo 

 & 0 & O & 29.53 & -250.19 & -28670.47 \\
 & 3 & O & 48.54 & 15.40 & -3351.46 \\
 & 3 & - & 75.34 & 29.53 & -4624.66\\
\hline
GPT-4-turbo 
 & 0 & O & 36.74 & -196.20 & -23863.26 \\
 & 3 & O & 67.84 & -18.91 & -8832.16 \\
 & 3 & - & 85.87 & 42.98 & -4314.13\\
\hline
GPT-4 
 & 0 & O & 38.40 & -215.98 & -26061.60\\
 & 3 & O & 79.82 & 15.50 & -6520.18\\
 & 3 & - & 90.45  & 62.18 & -2809.55 \\
\hline
% Finetuned-GPT & - & O & 98.05 & 94.64 & 91.23 & -601.95 \\
Finetuned-GPT & - & O & \textbf{98.05} & \textbf{91.23} & \textbf{-601.95}\\
\hline
\end{tabular}
}% End resizebox
\caption{Training set performance. Comparison of GPT models.}
\label{tab:train_fold_one}
\end{table}

\begin{table}[t]
\centering
\setlength{\tabcolsep}{10pt}
\renewcommand{\arraystretch}{1.2}
\resizebox{\columnwidth}{!}{
\begin{tabular}{lccc} \hline

\textbf{Model} & \textbf{RS(0)} & \textbf{RS(10)} & \textbf{RS(N)} \\ \hline\hline

GPT-3.5-turbo 
 & 70.34  & 13.59 & -6529.66 \\

GPT-4-turbo 

 &  76.53 & -6.02 & -9523.47 \\

GPT-4 

 &  79.28  & -17.88 & -11220.72 \\
\hline
Finetuned-GPT 
& \textbf{93.12}  & 50.99 & -4806.88 \\
\quad w/ table info. in SQL form
&83.23 &  17.02 & -7616.77 \\
\quad w/ Self-Reflection
&  83.15 & 62.51 & -2316.85 \\
\quad w/ Ensemble
&  91.57 & \textbf{74.38} & \textbf{-1908.43} \\
\hline
\end{tabular}
}
\caption{Ablation study conducted on the development set showcases the performance of in-context learning with few examples using GPT-3.5-turbo, GPT-4-turbo, and GPT-4, alongside fine-tuning performed with GPT-3.5-turbo using various additional techniques.
}
\label{tab:dev_score}
\end{table}

\subsection{In-Context Learning and Fine-tuning}
To evaluate the effectiveness of various GPT models for Text-to-SQL tasks, we conduct experiments with GPT-3.5-turbo, GPT-4-turbo, and GPT-4 for in-context learning and a fine-tuned version of GPT-3.5-turbo for supervised fine-tuning. Due to submission limitations, we assessed the GPT models using the training set. We adopt a k-fold cross-validation method with \(k=5\), training on four folds and evaluating the remaining fold. To maintain the balance of answerable and unanswerable questions in the training dataset, we divide unanswerable questions into three categories. When partitioning the training dataset into five folds, we ensured that the proportions of these categories were reflected in each fold. A detailed analysis of these categorized groups can be found in section 5.2. 

Table ~\ref{tab:train_fold_one} presents the comparison results of the GPT models. We experimented with variations by providing few-shot examples and including table information. The fine-tuned GPT model demonstrates superior performance across all metrics, making it our model of choice. Interestingly, the inclusion of table information slightly reduces performance in all in-context learning scenarios. We speculate that the table information in our experiment, which merely lists table names and column names, lacks detailed relational data like primary and foreign keys. Consequently, this minimal and potentially uninformative text might have acted as a distraction.

\subsection{Table Information Format}
The prompt includes table information to accurately identify the table and column names. Following the DAIL-SQL approach~\citep{Gao2023TexttoSQLEB}, we explore different formats of presenting table information, in both natural language and SQL format, within the same prompt framework. Our experiments, detailed in table~\ref{tab:dev_score}, reveal that presenting table information in SQL format results in a decrease in the RS (10) score from 50.99 to 17.02. This suggests that natural language formats are more readily interpretable by the language model such as GPT.

% Table selection performance comparison

\begin{table}[t]
\centering
\renewcommand{\arraystretch}{1.2}
\resizebox{\columnwidth}{!}{
\begin{tabular}{lccc}
\hline

\\ \hline
Model & Inclusion & Jaccard & Exact Match \\ \hline\hline
GPT-3.5-turbo & 0.7933 & 0.7930 & 0.7836 \\ 
GPT-4-turbo & 0.8912 & 0.8908 & 0.8723 \\
GPT-4 & 0.9250 & 0.9244 & 0.9123 \\
\textbf{Finetuned-GPT} & \textbf{0.9857} & \textbf{0.9855} & \textbf{0.9844} \\ 
Table Selector (GPT-3.5-turbo) & 0.8976 & 0.8488 & 0.7115 \\ \hline
\end{tabular}
}
\caption{Table selection performance.}
\label{tab:table_sel}
\end{table}

\subsection{Table Selection Results}
Considering the complexity of over ten tables and the resulting SQL queries that reference multiple tables, we hypothesize that a self-reflection incorporating selective, detailed table information could enhance the accuracy of the generated SQL queries. In preparation for this self-reflection process, we assess the accuracy of the tables retrieved in the generated SQL queries. This assessment involves calculating the accuracy between correct tables and the tables extracted from the generated SQL queries. We use three metrics as an accuracy score: 1) inclusion score (indicating the presence of the correct tables within the generated SQL), 2) the Jaccard similarity score (comparing the intersection to the union of correct and extracted tables), 3) and the exact match score. 

Table ~\ref{tab:table_sel} suggests that the fine-tuned GPT model effectively identifies the relevant tables without a dedicated table selection model. We extract tables from the initially generated SQL queries and use prompts augmented with detailed information, such as descriptions of each column and examples of values, for the fine-tuned GPT model. The comparison between the initially generated SQL and the outcomes after the self-reflection stage, table~\ref{tab:dev_score} shows an increase in the RS(10) score from 50.99 to 62.51 in the development set, and table~\ref{tab:test_score} also illustrates an improvement in the RS(10) score from 13.80 to 27.85 in the test set. This improvement indicates that the regenerated SQL queries provide more reliable and accurate outputs.

\begin{table}[t]
\centering
\setlength{\tabcolsep}{10pt}
\renewcommand{\arraystretch}{1.2}
\resizebox{\columnwidth}{!}{
\begin{tabular}{lccc} \hline
% \rowcolor{Gray}
\textbf{Model} & \textbf{RS(0)} & \textbf{RS(10)} & \textbf{RS(N)} \\ \hline\hline

Finetuned-GPT (Ensemble)
& 72.07 & \textbf{59.21} & \textbf{-1427.93} \\ \hline
Finetuned-GPT (Single)
& 78.06  & 13.80 & -7421.94 \\
\quad w/ Self-Reflection
&  77.55 & 27.85 & -5722.45\\
\quad   w/ Question Templatization
&  \textbf{80.55} & 40.27 &-4619.45\\

\hline
\end{tabular}
}
\caption{Ablation study on the test set. We provide the performance of ensembled and single results. Every component, including SQL regeneration and question templatization, plays a key role in enhancing overall performance.
}
\label{tab:test_score}
\end{table}

\subsection{Question Templatization}
Our analysis focuses on the characteristics of the questions across each dataset. It reveals a decline in the fine-tuned GPT model's scores from the training set to the development and test sets. This pattern highlights substantial variations among the training, development, and test datasets. To mitigate these discrepancies, we employ the technique that reverses the deviation of questions from templates. We utilize GPT-4-turbo to rephrase the original question. By prompting GPT-4-turbo with the original question and semantically similar questions from the training set and template from \citep{lee2022ehrsql}, we aim to achieve consistency with related queries. This approach significantly reduces the distribution gaps between the training and test sets, as demonstrated in Table~\ref{tab:test_score}. The improvement in the RS(10) score from 13.80 to 40.27 highlights the effectiveness of question templatization by comparing the performance of a single model before and after its application.

\section{Analysis}
In this section, we analyze the word distribution of questions for each dataset split: training, development, and test sets. The objective is to identify variations in question composition among these datasets. Furthermore, we investigate the distribution of unanswerable questions in the training set to better understand questions that yield an \textit{null} response.

To focus solely on word analysis and minimize noise, we eliminate punctuation marks such as ".", ",", and "?", remove stop words such as "the", "a", and "an" from the questions, and convert all letters to lowercase. After eliminating these elements, we analyze the processed questions using N-grams. The analysis is limited to 1 to 3-grams, which is sufficient for understanding the context of questions while excluding the aforementioned noise. ~\ref{app:word_freq} details the ten most prevalent words alongside their respective frequencies within each dataset arranged in non-increasing order, including the collection of unanswerable queries labeled as \textit{Unanswerable Train set}.

\subsection{N-gram Distribution}
The initial three columns of~\ref{app:word_freq} enumerates the top ten most frequent words in each dataset alongside their respective frequencies. Analysis of~\ref{app:word_freq} indicated that words with high frequency within one dataset tended to be frequent across other datasets as well, suggesting a pattern of similarity. However, it was observed that words with lower frequency, which were not included in the table, often did not appear in other datasets. This discrepancy became particularly evident within the context of 3-gram sets, highlighting a distinct distribution among the datasets.

This disparity underscores the necessity of developing a robust model that can adapt and excel across datasets with different word distributions.

\subsection{Category of Unanswerable Questions}

We analyzed the training set's \textit{null} distribution, identifying 450 unanswerable questions. Our initial qualitative analysis involved categorizing these \textit{null}-labeled questions into three distinct groups through a detailed manual review: (1) Incorrect Patient Number, (2) Require External Knowledge, (3) Out of EHR Database.

In the first case, based on the MIMIC-IV dataset's criteria, a legitimate patient number is identified by its 8-digit configuration; thus, questions featuring a patient number with fewer or more than 8 digits invariably resulted in a \textit{null} response. Regarding the second case, specific questions, for example, \textit{"I am curious what the protocols for the drugs that work to treat cancer."} could potentially be answered by a knowledgeable individual or through QA tasks using external information resources. The third group, while seemingly akin to the second, differed in that the questions could technically be converted into SQL queries; however, they remained unanswerable due to the absence of the required information in the dataset. Example questions include: \textit{"Has patient 23224 an appointment in another hospital department?"}. Further examples for each category, along with their respective frequencies, are detailed in~\ref{app:unans_examples}.

Additionally, a quantitative analysis of unanswerable questions was also conducted using N-grams. By examining the differences in word distribution between "answerable" and "unanswerable" questions, as highlighted by the contrast between the first and last columns of~\ref{app:word_freq}, significant disparities were noted. For instance, an examination of the 1-gram columns for both the training set and the Unanswerable Training set reveals that the only overlapping words are "patient" and "last." This indicates a significantly different distribution between the two datasets. 

Based on both qualitative and quantitative analysis, we were able to refine \method\ to avoid generating SQL queries for questions that solely comprise words found in the unanswerable questions of the training set.

\section{Conclusion}
Throughout the challenge, we noticed that differences in the way data is distributed across training, development, and test sets can make it hard for our model to determine which questions are answerable or not. To tackle this issue, we templatized questions to make the word distribution of development and test data more similar to the training data. This method aimed to bridge the gap between the datasets, helping the model better understand the features of unanswerable questions within the test dataset.

Although we did not address this in this study, we anticipate that future research could see performance improvements by augmenting the training dataset to more closely match the distribution of unanswerable questions in the development and test sets. Focusing on refining the test data to align more closely with the characteristics observed in the training datasets, we expect to increase model performance in identifying unanswerable questions. Such data augmentation strategies could bridge the remaining gaps between datasets and ensure a more robust model performance across varied datasets. Also, we utilized finetuned gpt-3.5-turbo, which is expensive and unusable for other researchers. Thus, further study should be done with open sourced models, like llama or gemma. 

\section*{Acknowledgments}

This work was supported in part by the National Research Foundation
of Korea [NRF2023R1A2C3004176], the Ministry of Health \& Welfare,
Republic of Korea [HR20C0021(3)], the Ministry of Science and ICT
(MSIT) [RS-2023-00262002].
\bibliography{EHR2SQL}

\clearpage
\onecolumn
\appendix
\renewcommand{\thesection}{Appendix \Alph{section}}

\section{Examples of unanswerable questions}
\label{app:unans_examples}
\begin{table}[H]
\centering
\small
\renewcommand{\arraystretch}{1.5}

% \begin{table}[H]
\centering
\scalebox{0.95}{
\begin{tabular}{|p{0.15\textwidth}|p{0.7\textwidth}|p{0.15\textwidth}|}
\hline
\textbf{Category} & \textbf{Example} & \textbf{Frequency}\\
\hline
\multirow{3}{=}{Incorrect Patient Number} & Will they have any urine test done for patient 24628? & \multirow{3}{*}{252 (56\%)}\\
\cline{2-2}
& Is patient 21074 subject to tests involving covid-19? & \\
\cline{2-2}
& Do you know what type of blood patient 1903 has? &\\
\hline
\multirow{3}{=}{Require External Knowledge} & What is a checklist before lumb/lmbosac fus ant/ant? & \multirow{3}{*}{83 (18.4\%)} \\
\cline{2-2}
& What is the protocol used for the anticancer drugs? & \\
\cline{2-2}
& So tell me what to do before you go for hemodialysis. & \\
\hline
\multirow{3}{=}{Out of EHR Knowledge Base} & What kind of blood patient 18866 has. & \multirow{3}{*}{115 (25.6\%)} \\
\cline{2-2}
& List the single rooms that are available now? &\\
\cline{2-2}
& When are dr. oneill's rounds and procedures? &\\
\hline
\end{tabular}
}
\caption[Examples of Unanswerable Questions with Respective Frequencies]{\centering Examples of Unanswerable Questions with Respective Frequencies}
\label{tab:appendix_unans_ex}
% \end{table}

\end{table}

\clearpage
\section{Prompt}
\label{app:prompt}
\begin{table}[H]
\centering
\includegraphics[width=0.9\textwidth]{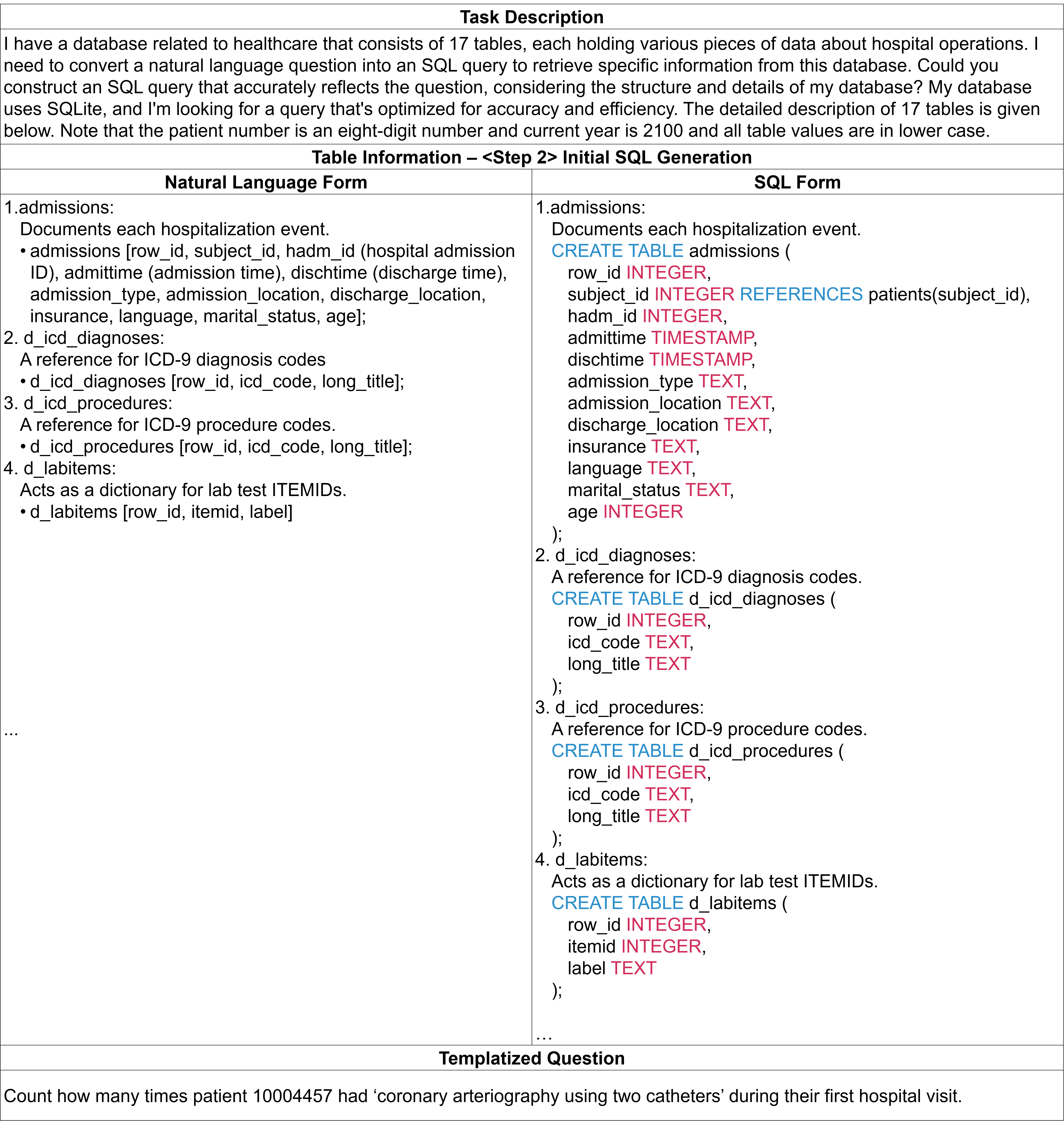}
\caption{The prompt used in the step 2 initial SQL generation.}
\label{tab:step2}
\end{table}

\clearpage
\section{Detailed Prompt Example}
\label{app:example_appendix}
\begin{table}[H]
\centering
\includegraphics[width=0.9\textwidth]{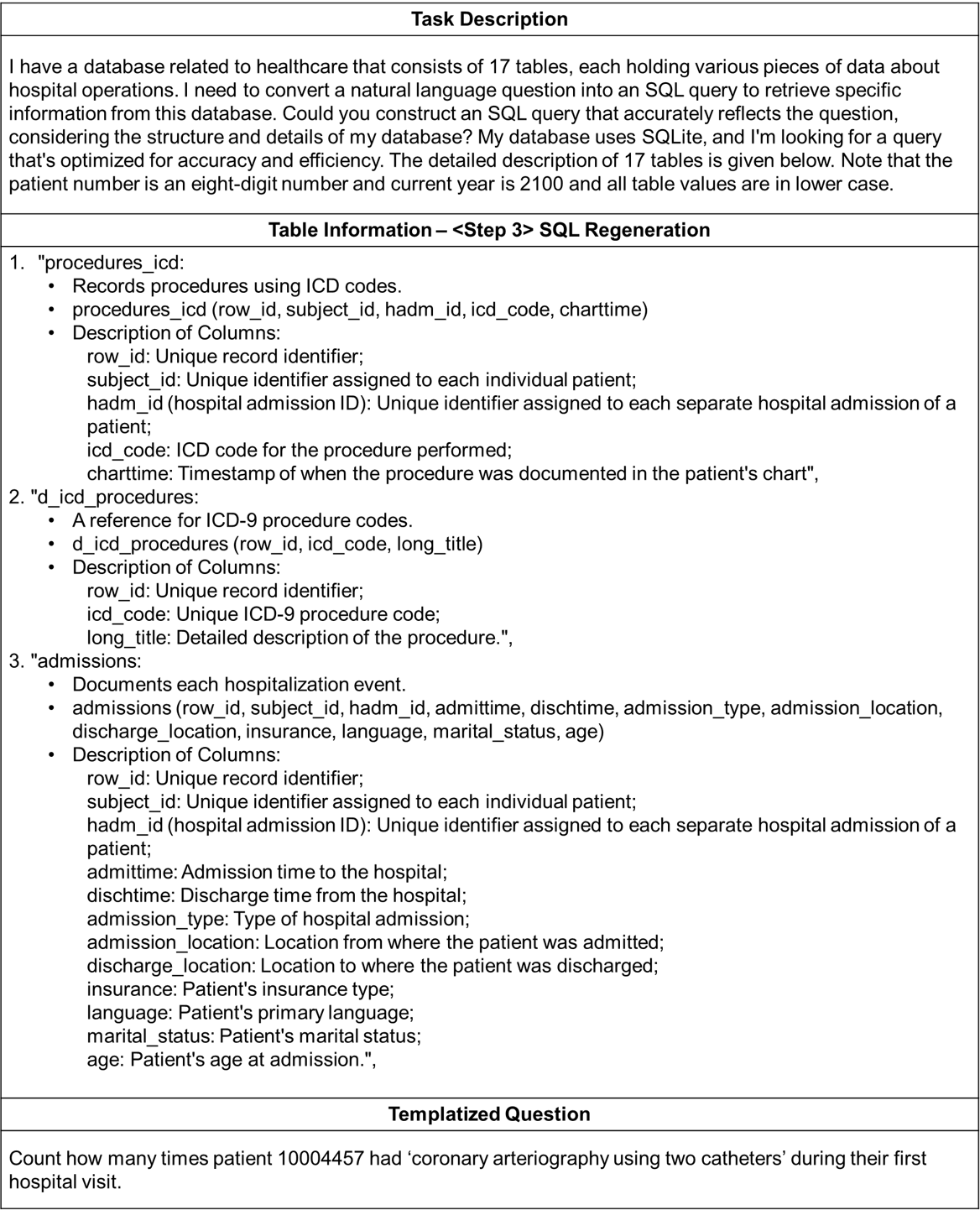}
\caption{The prompt used in Step 3 for SQL regeneration.}
\label{tab:step3}
\end{table}

\clearpage
\section{Word frequencies: 3-gram}
\label{app:word_freq}
\begin{table}[H]
\centering
\small
\renewcommand{\arraystretch}{1.2}

\centering
\begin{tabular}{|l|l|l|l|}
\hline
N-gram &
  \textbf{1-gram} &
  \textbf{2-gram} &
  \textbf{3-gram} \\ \hline
\textbf{Train set} &
  \begin{tabular}[c]{@{}l@{}}('patient',): 3205\\ ('since',): 1572\\ ('last',): 1394\\ ('hospital',): 1340\\ ('first',): 1232\\ ('year',): 934\\ ('patients',): 861\\ ('2100',): 818\\ ('visit',): 778\\ ('time',): 734\end{tabular} &
  \begin{tabular}[c]{@{}l@{}}('hospital', 'visit'): 608\\ ('since', '2100'): 425\\ ('first', 'hospital'): 327\\ ('last', 'time'): 317\\ ('hospital', 'encounter'): 316\\ ('last', 'hospital'): 302\\ ('since', '1'): 302\\ ('1', 'year'): 298\\ ('year', 'ago'): 298\\ ('first', 'time'): 280\end{tabular} &
  \begin{tabular}[c]{@{}l@{}}('since', '1', 'year'): 298\\ ('1', 'year', 'ago'): 298\\ ('first', 'hospital', 'visit'): 203\\ ('last', 'hospital', 'visit'): 183\\ ('within', '2', 'months'): 153\\ ('last', 'time', 'patient'): 118\\ ('first', 'time', 'patient'): 104\\ ('first', 'hospital', 'encounter'): 94\\ ('last', 'hospital', 'encounter'): 92\\ ('measured', 'last', 'hospital'): 92\end{tabular} \\ \hline
\textbf{Dev set} &
  \begin{tabular}[c]{@{}l@{}}('patient',): 656\\ ('hospital',): 322\\ ('since',): 316\\ ('patients',): 279\\ ('last',): 255\\ ('first',): 253\\ ('year',): 202\\ ('visit',): 199\\ ('2100',): 170\\ ('time',): 134\end{tabular} &
  \begin{tabular}[c]{@{}l@{}}('hospital', 'visit'): 150\\ ('since', '2100'): 82\\ ('last', 'hospital'): 72\\ ('since', '1'): 70\\ ('1', 'year'): 69\\ ('year', 'ago'): 69\\ ('first', 'time'): 67\\ ('hospital', 'encounter'): 65\\ ('first', 'hospital'): 59\\ ('lab', 'test'): 46\end{tabular} &
  \begin{tabular}[c]{@{}l@{}}('since', '1', 'year'): 69\\ ('1', 'year', 'ago'): 69\\ ('last', 'hospital', 'visit'): 46\\ ('first', 'hospital', 'visit'): 40\\ ('within', '2', 'months'): 30\\ ('last', 'hospital', 'encounter'): 20\\ ('first', 'time', 'patient'): 20\\ ('current', 'hospital', 'visit'): 19\\ ('arterial', 'blood', 'pressure'): 19\\ ('top', 'three', 'frequent'): 18\end{tabular} \\ \hline
\textbf{Test set} &
  \begin{tabular}[c]{@{}l@{}}('patient',): 620\\ ('hospital',): 318\\ ('patients',): 306\\ ('since',): 304\\ ('last',): 265\\ ('first',): 234\\ ('year',): 217\\ ('2100',): 184\\ ('visit',): 164\\ ('prescribed',): 140\end{tabular} &
  \begin{tabular}[c]{@{}l@{}}('hospital', 'visit'): 128\\ ('since', '2100'): 100\\ ('hospital', 'encounter'): 81\\ ('first', 'hospital'): 72\\ ('last', 'hospital'): 72\\ ('since', '1'): 57\\ ('1', 'year'): 57\\ ('year', 'ago'): 57\\ ('many', 'patients'): 54\\ ('number', 'patients'): 47\end{tabular} &
  \begin{tabular}[c]{@{}l@{}}('since', '1', 'year'): 57\\ ('1', 'year', 'ago'): 57\\ ('first', 'hospital', 'visit'): 42\\ ('last', 'hospital', 'visit'): 42\\ ('within', '2', 'months'): 41\\ ('last', 'hospital', 'encounter'): 26\\ ('measured', 'last', 'hospital'): 23\\ ('first', 'time', 'patient'): 20\\ ('first', 'hospital', 'encounter'): 20\\ ('last', 'time', 'patient'): 18\end{tabular} \\ \hline
\textbf{\begin{tabular}[c]{@{}l@{}}Unanswerable \\Train set\end{tabular}} &
  \begin{tabular}[c]{@{}l@{}}('patient',): 252\\ ('department',): 49\\ ('tell',): 42\\ ('procedure',): 41\\ ('blood',): 36\\ ('dr',): 36\\ ('received',): 34\\ ('rooms',): 29\\ ('test',): 28\\ ('last',): 27\end{tabular} &
  \begin{tabular}[c]{@{}l@{}}('received', 'department'): 20\\ ('outpatient', 'schedule'): 18\\ ('rounds', 'procedures'): 17\\ ('another', 'department'): 16\\ ('rooms', 'available'): 15\\ ('diagnosis', 'patient'): 15\\ ('operating', 'rooms'): 14\\ ('blood', 'transfusion'): 14\\ ('name', 'diagnosis'): 14\\ ('genetic', 'test'): 14\end{tabular} &
  \begin{tabular}[c]{@{}l@{}}('name', 'diagnosis', 'patient'): 12\\ ('last', 'time', 'patient'): 11\\ ('many', 'operating', 'rooms'): 11\\ ('appointment', 'another', 'department'): 11\\ ('genetic', 'test', 'patient'): 10\\ ('subject', 'covid-19', 'testing'): 9\\ ('type', 'blood', 'patient'): 9\\ ('ward', 'id', 'patient'): 9\\ ("today's", 'outpatient', 'schedule'): 9\\ ('outpatient', 'schedule', 'dr'): 8\end{tabular} \\ \hline
\end{tabular}

\caption{3-Gram frequency table with 10 examples sorted in non-increasing order}
\label{tab:three-gram}

\end{table}

\clearpage
\section{Examples of Question Templating}
\label{app:question_templatization}
\begin{table}[H]
\centering
\renewcommand{\arraystretch}{1.5}

\begin{table}[H]
\centering
\small
\begin{tabular}{|m{0.2\textwidth}|m{0.75\textwidth}|}
\hline
\multicolumn{2}{|l|}{\textbf{Input: "When does patient 8016's influenza quarantine end?"}} \\ \hline
\multirow{3}{*}{\centering \textbf{Candidate templates}} 
& \parbox[t]{\hsize}{\strut "What was the time of <patient number>'s last influenza a/b by dfa microbiology test since 03/2100?" \strut} \\ \cline{2-2}
& \parbox[t]{\hsize}{\strut "Can you tell me when <patient number> had their first rapid respiratory viral screen \& culture microbiology test in 08/this year?" \strut} \\ \cline{2-2}
& \parbox[t]{\hsize}{\strut "When did <patient number> depart hospital during this year for the last time?" \strut} \\ \hline
\multicolumn{2}{|l|}{\textbf{Reformulated: "When is the end date of patient 8016's influenza quarantine?"}} \\ \hline

\multicolumn{2}{|l|}{\textbf{Input: "Pull up the IDs of patients who were diagnosed with cataract extraction status."}} \\ \hline
\multirow{3}{*}{\centering \textbf{Candidate templates}} 
& \parbox[t]{\hsize}{\strut "Number of patients who were diagnosed with <unk>." \strut} \\ \cline{2-2}
& \parbox[t]{\hsize}{\strut "Number of patients who were diagnosed throughout this year with <unk>." \strut} \\ \cline{2-2}
& \parbox[t]{\hsize}{\strut "Tell me the number of patients diagnosed with <unk>." \strut} \\ \hline
\multicolumn{2}{|l|}{\textbf{Reformulated: "Retrieve the ids of patients diagnosed with 'cataract extraction status.'"}} \\ \hline

\multicolumn{2}{|l|}{\textbf{Input: "How many duloxetine prescription cases were there since 1 year ago?"}} \\ \hline
\multirow{3}{*}{\centering \textbf{Candidate templates}} 
& \parbox[t]{\hsize}{\strut "How much duloxetine has been prescribed to <patient number> in 05/2100 in total?" \strut} \\ \cline{2-2}
& \parbox[t]{\hsize}{\strut "How many drugs have been prescribed to <patient number> since 2 months ago?" \strut} \\ \cline{2-2}
& \parbox[t]{\hsize}{\strut "What is the number of drugs <patient number> was prescribed since 1 year ago?" \strut} \\ \hline
\multicolumn{2}{|l|}{\textbf{Reformulated: "What is the number of duloxetine prescription cases since 1 year ago?"}} \\ \hline
\end{tabular}
\caption{Examples of question templating using question masked templates.}
\label{tab:example_questions}
\end{table}

\label{tab:re_query_ex}
\end{table}

\clearpage
\twocolumn

\end{document}